\documentstyle[twoside,fleqn,espcrc2,epsfig]{article}

\newcommand{\chup}{\chi U\phi}

\newcommand{\chb}{\overline{\chi}}
\newcommand{\Reop}{\mathop{\rm Re}\nolimits}

\textfloatsep=\floatsep          
\intextsep=\floatsep             

\title{%
  \mbox{}\hbox to 0pt{\vbox to 0pt{\vss
      \parbox[b]{\hsize}{\normalsize
        \begin{flushright}
          HLRZ 51/96\\
          hep-lat/9607036
        \end{flushright}
        \vspace*{2.5cm}}}\hss}%
  Scaling behavior at the tricritical point in the fermion-gauge-scalar
  model\thanks{Work supported by DFG and BMBF. The computations have been
    performed on VPP500 of RWTH Aachen and CRAY-YMP/T90 of HLRZ J\"ulich.}}

\author{W. Franzki\address{Institute of Theoretical Physics E, RWTH
    Aachen, D-52056 Aachen, Germany\\ and HLRZ c/o Forschungszentrum KFA,
    D-52425 J{\"u}lich, Germany}%
    \thanks{Work done in collaboration with J. Jers\'ak.}%
  }

\begin{document}

\begin{abstract}
  We investigate a strongly coupled U(1) gauge theory with fermions and
  scalars on the lattice and analyze whether the continuum limit might be a
  renormalizable theory with dynamical mass generation.  Most attention is
  paid to the phase with broken chiral symmetry in the vicinity of the
  tricritical point found in the model. There we investigate the scaling of
  the masses of the composite fermion and of some bosonic bound states.  As a
  by-product we confirm the mean-field exponents at the endpoint in the
  U(1)-Higgs model, by analyzing the scaling of the Fisher zeros.
\end{abstract}

\maketitle

\section{Introduction}
The fermion-gauge-scalar-model ($\chi U\phi$ model) was suggested as a model
for dynamical mass generation in \cite{FrJe95a}. It has a confining phase with
dynamical chiral symmetry breaking (Nambu phase).  The physical fermion is a
bound-state of the fundamental fermion and the scalar: $F=\phi^\dagger\chi$.
It is neutral under the (strong) gauge interaction and thus escapes the
confinement. Nevertheless, it acquires a mass, which scales to zero at a
2$^{\rm nd}$ order phase transition (PT)~\cite{FrFr95a,FrJe96a}.  These
features suggest the existence of a new mass generation mechanism,
different from the standard Higgs-Yukawa sector.

\section{The model}
The model is defined by the lattice action
\begin{equation}
  S_{\chi U \phi} = S_\chi + S_U + S_\phi \; ,
  \label{action}
\end{equation}
where
\begin{eqnarray}
  S_\chi \hspace{-2mm}&=&\hspace{-2mm} {\textstyle \frac{1}{2}}
  \sum_x \sum_{\mu = 1}^4
  \eta_{\mu x} \chb_x \left[ U_{x,\mu} \chi_{x + \mu} -
    U_{x-\mu,\mu}^\dagger  \chi_{x-\mu} \right]\!
  \nonumber\\
  && + \, a m_0 \sum_x \chb_x \chi_x \; ,
\label{SCH}                  \\
  S_U \hspace{-2mm}&=&\hspace{-2mm} \beta \, \sum_{\rm P} \left[ 1 - \Reop
    U_{\rm P} \right] \; ,
\label{SU}  \\
  S_\phi \hspace{-2mm}&=&\hspace{-2mm} - \kappa \, \sum_x \sum_{\mu=1}^4 \left[
    \phi_x^\dagger U_{x,\mu} \phi_{x + \mu} \,+\, {\rm h.c.} \right] \; .
\label{SPH}
\end{eqnarray}
Here $U_{\rm P}$ is the plaquette product of link variables $U_{x,\mu}$ and
$\eta_{\mu x} = (-1)^{x_1 + \cdots + x_{\mu - 1}}$. The gauge field link
variables $U_{x,\mu}$ are elements of the compact gauge group U(1).  The
complex scalar field $\phi$ of unit charge satisfies the
constraint~$|\phi|=1$.
The staggered fermion field $\chi$ of charge one leads to the global U(1)
chiral symmetry of the model in the chiral limit, i.\,e.\ when the bare fermion
mass $am_0$ vanishes.

A detailed discussion of the phase diagram and its investigation can be found
in \cite{FrFr95a,FrJe96a}.
 
\section{Scaling in the quenched approximation}
A major tool in the examination of universality classes is the determination
of critical exponents by finite size scaling analysis. To get experience with
this method we investigate the model in the quenched approximation, which
corresponds to $am_0=\infty$.  This model is known by itself under the names
`scalar QED' or `U(1)-Higgs model'. In this model the 1$^{\rm st}$ order
Higgs-PT ends in a critical point (E$_\infty$).  The scaling at this point was
investigated in \cite{AlAz93} along the 1$^{\rm st}$ order line, indicating
mean-field exponents.  We investigate the scaling along different lines
passing through E$_\infty$.

\begin{figure}[tbp]
  \vspace*{-7mm}%
  \begin{center}
    \leavevmode
    \hspace*{-8mm}%
    \epsfig{file=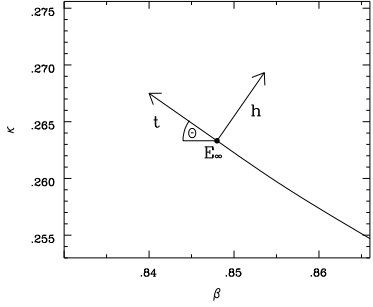,height=7cm,angle=90}\hspace*{-18mm}%
    \vspace*{-13mm}%
    \caption{Diagram of the new parameters $t$ and $h$. The point
    E$_\infty$ is the endpoint of the Higgs-PT line (indicated
    bold).}
    \label{fig:par}
  \end{center}
\end{figure}

First of all it is convenient to introduce two reduced couplings
(fig.~\ref{fig:par}):\\
\begin{tabular}[t]{@{}ccl}
  $t$ &:& parallel to the 1$^{\rm st}$ order PT line and\\
  $h$ &:& perpendicular to the PT line.
\end{tabular}\\
The letter $t$ and $h$ have been chosen in correspondence to temperature and
external field of a magnetic system.

We now introduce critical exponents for both directions.  The correlation
length critical exponents $\nu$ and $\tilde\nu$ are defined by
\begin{eqnarray}
  \xi &\propto& |t|^{-\nu}|_{h=0} \mbox{ \ and}\\
  \xi &\propto& |h|^{-\tilde{\nu}}|_{t=0}\;,
\end{eqnarray}
with $\xi$ being the correlation length. The corresponding exponents connected
with the heat capacity are called $\alpha$ and $\tilde\alpha$.

To understand the connection between $\nu$ and $\tilde\nu$ we regard the
scaling relation
\begin{equation}
  \xi = |t|^{-\nu} F\left(\frac{|h|}{|t|^\Delta}\right)
\end{equation}
with the scaling function $F$ and $\Delta = \beta+\gamma$~\cite{Hu87}.
This transforms with $\tilde{F}(x)=  x^{\nu} F(x^\Delta)$ to
\begin{equation}
  \xi = |h|^{-\nu/\Delta}
  \tilde{F}\left(\frac{|h|^{1/\Delta}}{|t|}\right)\;.
\end{equation}
Assuming $\tilde{F}(\infty)<\infty$ this means $\tilde\nu=\nu/\Delta$.


The mean-field exponents $\beta=1/2$, $\gamma=1$ and $\nu=1/2$ correspond
to $\tilde{\nu}=1/3$ and $\tilde\alpha/\tilde\nu=2$.

Typically, the most precise way to determine the correlation length critical
exponent numerically is the measurement of the finite size scaling of the edge
singularity in the complex coupling plane (Fisher zeros). From scaling
arguments for the free energy we expect for the first zero $z_0$:
\begin{equation}
  \left.\mathop{\rm Im}\nolimits z_0(L)\right|_{t=0} = A \cdot L^{-1/\tilde{\nu}}\;.
\end{equation}
Because all directions non-parallel to the 1$^{\rm st}$ order PT
are equivalent, we expect the same exponent $\tilde\nu$ also if we
fix $\beta=\beta_{\rm E_\infty}$ or $\kappa=\kappa_{\rm E_\infty}$.

We did a numerical simulation at $\beta=0.848\simeq \beta_{\rm E_\infty}$ on
lattices from $4^4$ to $16^4$ at 5 to 10 $\kappa$ values with a statistic per
point between 32000 and 192000 measurements, each separated by 4 Metropolis
sweeps.  For the determination of the zeros we use a multihistogram
reweighting with 10000 bins.

\begin{figure}[tbp]
  \vspace*{-3mm}%
  \begin{center}
    \leavevmode
    \hspace*{-8mm}%
    \epsfig{file=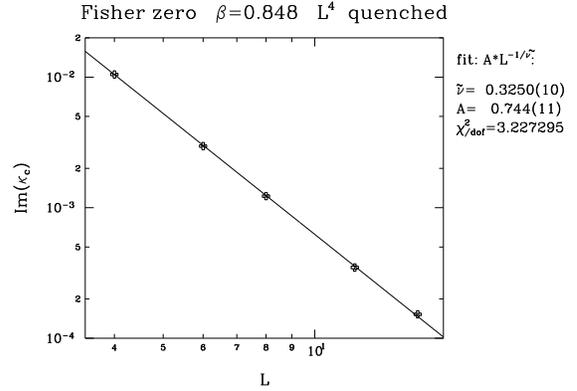,height=8.2cm,angle=90}\hspace*{-5mm}%
    \vspace*{-13mm}%
    \caption{Scaling of the edge singularity for $\beta=0.848\simeq \beta_{\rm
        E_\infty}$. The value for $\beta_{\rm E_\infty}$ has been taken from
      \protect\cite{AlAz93}.}
    \label{fig:fisher1}
  \end{center}
\end{figure}

Fig.~\ref{fig:fisher1} shows nice scaling for
all lattice sizes with a critical exponent $\tilde\nu=0.3250(10)$.
This exponent is quite near to the mean-field exponent $\tilde\nu=1/3$.  The
small deviations outside the error bars might be due to logarithmic
corrections, which are to be expected at a Gaussian fixpoint. We intend to
check this by methods as described in \cite{KeLa93} for the 4d Ising model.

A similar exponent has also been measured in the SU(2)-Higgs
model~\cite{BoEv90b,Bo90}, but it was not realized, that this is compatible
with a Gaussian fixpoint.

To check this scaling, we have also determined the specific heat and a fourth
order cumulant:
\begin{eqnarray}
  \label{defcum}
  c_V &=& \frac{1}{6V}\left\langle(E-\langle E\rangle)^2\right\rangle\;,\\
  V_{\rm CLB} &=& -\frac{1}{3}\frac{\left\langle(E^2-\langle
      E^2\rangle)^2\right\rangle}{\langle E^2\rangle^2}\;.
\end{eqnarray}
Here $E$ can be any linear combination of $E_{\rm L}$ and $E_{\rm P}$ thats
picks up also the orthogonal component $E_\perp$ (for def. of $E_\perp$ see
\cite{FrFr95a}).  We checked this for $E_{\rm L}$, $E_{\rm P}$ and $E_\perp$.

\begin{figure}[tbp]
  \vspace*{-3mm}%
  \begin{center}
    \leavevmode
    \hspace*{-8mm}%
    \epsfig{file=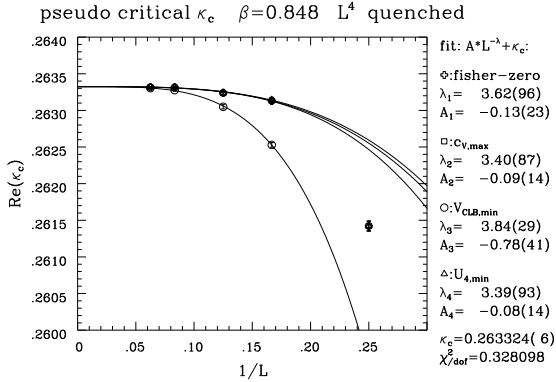,height=8cm,angle=90}\hspace*{-5mm}%
    \vspace*{-13mm}%
    \caption{Scaling of the pseudocritical couplings with lattice size. All
      data have been fitted with one $\kappa_c$. Only lattices with $L\leq 6$
      have been considered in the fit.}
    \label{fig:kcre}
  \end{center}
\end{figure}

We expect the following scaling relation to hold:
\begin{eqnarray}
  \label{scalcum}
  c_{V,\it max}(L) &=& A \cdot L^{\tilde{\alpha}/\tilde{\nu}}\;,\\
  V_{{\rm CLB},\it min}(L) &=& A \cdot L^{\tilde{\alpha}/\tilde{\nu}-4}\;.
\end{eqnarray}
There might be a regular contribution to $c_{V,\it max}$, but our data show no
indication for this, and so we do not add any constant in our fits.

The first two observables show nice scaling for all lattice sizes with
exponents in good agreement of those determined by the Fisher zeros. Small
deviations indicate systematic uncertainties a little larger than the
statistical errors.


We also checked the finite size scaling of the real part of the fisher zeros,
as we do for the pseu\-do\-cri\-ti\-cal coupling, determined by the extrema of
the cumulants (fig.~\ref{fig:kcre}). The value of the shift exponent $\lambda$
is compatible with $1/\tilde\nu\simeq 3$.

\begin{figure}[tbp]
  \vspace*{-3mm}%
  \begin{center}
    \leavevmode
    \hspace*{-8mm}%
    \epsfig{file=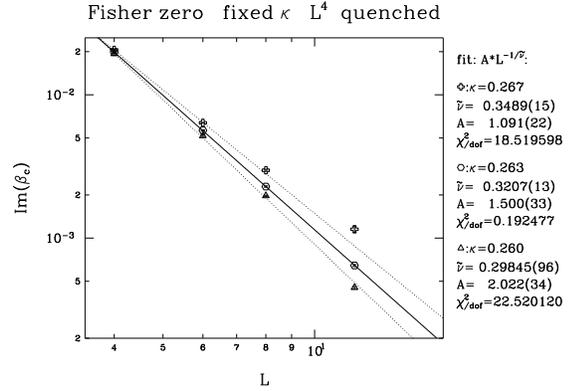,height=8.2cm,angle=90}\hspace*{-5mm}%
    \vspace*{-12mm}%
    \caption{Scaling of the edge singularity for three different $\kappa$. From
      top to bottom: $\kappa=0.267$, $\kappa=0.263\simeq\kappa_{\rm E_\infty}$ and
      $\kappa=0.26$.}
    \label{fig:fisher2}
  \end{center}
\end{figure}

To check the claim, that $\tilde\nu$ is independent of the direction, we
also made some runs for fixed $\kappa$. The results are shown in
fig.~\ref{fig:fisher2}.
The critical endpoint corresponds to $\kappa=0.263$. The critical exponent
$\tilde\nu=0.3206(15)$ is in good agreement with that for fixed
$\beta=0.848$.

In fig.~\ref{fig:fisher2} we also show the scaling for two $\kappa$ a little
bit off from the endpoint. $\kappa=0.267$ is beyond the endpoint, where no
critical behavior is expected. $\kappa=0.26$ is on the 1$^{\rm st}$ order
line, corresponding to $\tilde\nu=0.25$. In both cases clear deviations from a
linear scaling can be observed, with the right tendency for increasing lattice
size.  The straight lines are just plotted to visualize the deviations.

Both observations support the reliability of this finite size scaling method.


\section{Spectrum in the Nambu phase at the tricrital point E}

A rich spectrum is to be expected near the tricritical point E, significantly
different from that of the second order chiral PT line NE. This
is discussed in \cite{FrJe95a,FrJe96a} and here we want to show some new
results. We fix $\kappa=0.30\simeq\kappa_{\rm E}$, because the $\kappa$ of
the endpoint (for small fixed $am_0$) turned out to be less dependent on
$am_0$ than $\beta$.

As shown in \cite{FrJe96a} we observe strong finite size effects. The boson
mass $am_S$ (scalar bound state of $\phi$'s) shows a pronounced dip at the
critical point, which shifts with the volume and $am_0$. It turned out, that
those dips fall together, if one plots $am_S$ as a function of the
fermion mass $am_F$.

\begin{figure}[tbp]
  \vspace*{-4mm}%
  \begin{center}
    \leavevmode
    \hspace*{-8mm}%
    \epsfig{file=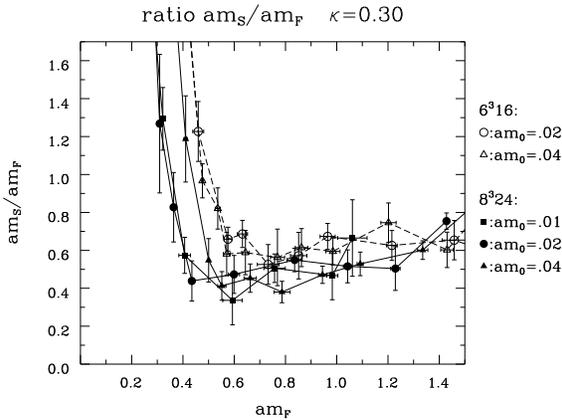,height=8.8cm,angle=90}\hspace*{-10mm}%
    \vspace*{-13mm}%
    \caption{Mass ratio $am_S/am_F$ as a function of $am_F$ for
      $\kappa=0.30\simeq\kappa_{\rm E}$.}
    \label{fig:massratio}
  \end{center}
\end{figure}

To investigate the scaling, we look for the mass ratio $am_S/am_F$
(fig.~\ref{fig:massratio}). The sudden increase of this ratio for small $am_F$
indicates the symmetric phase, with $am_F$ vanishing (up to finite size
effects). The nearly constant value of this ratio in the broken phase may
indicate that both particles survive in the continuum limit with a mass ratio
around 1/2.

\begin{figure}[tbp]
  \vspace*{-4mm}%
  \begin{center}
    \leavevmode
    \hspace*{-8mm}%
    \epsfig{file=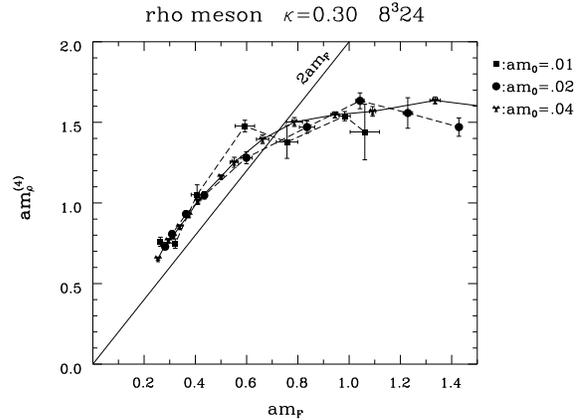,height=8.8cm,angle=90}\hspace*{-10mm}%
    \vspace*{-13mm}%
    \caption{The $\rho$-mass as a function of $am_F$.}
    \label{fig:mes}
  \end{center}
\end{figure}

We also measure different `mesons' (fermion-antifermion bound states), defined
in \cite{FrFr95a}. The mass of the $\sigma$ particle is very hard to measure,
due to its vacuum quantum numbers and the nonvanishing background expectation
value connected with this. So we show here the results for the $\rho$-meson
(fig.~\ref{fig:mes}).
We interpret this data as an indication for $\rho$ surviving in the continuum
limit as a resonance.

\section{Conclusions}
The investigation of the tricritical point in the $\chup_4$-model makes
further progress. The mass ratio $am_S/am_F$ approaches a constant value,
supporting the expectation, that the continuum limit is substantially
different from that of the NJL model. This strengthens the hope, that the
theory might be renormalizable and a model for dynamical mass generation. We
hope to get a deeper understanding of the scaling behavior by means of the
finite size scaling analysis.


As a by-product we confirm the observation in \cite{AlAz93}, that the
endpoint of the U(1)-Higgs model show mean-field like scaling, probably with
logarithmic corrections.

\bibliographystyle{wunsnot}   


\end{document}